\documentclass[12pt]{article}
\usepackage{a4wide}
\usepackage{amssymb}
\usepackage{graphicx}
\begin{document}
{\renewcommand{\thefootnote}{\fnsymbol{footnote}}
\hfill  IGPG--06/9--7\\
\medskip
\hfill gr--qc/0609057\\
\medskip
\begin{center}
{\LARGE Hamiltonian cosmological perturbation theory with loop quantum
gravity corrections}\\
\vspace{1.5em}
Martin Bojowald$^1$\footnote{e-mail address: {\tt
bojowald@gravity.psu.edu}}, Hector H.~Hern\'andez$^2$\footnote{e-mail
address: {\tt hehe@aei.mpg.de}}, Mikhail Kagan$^1$\footnote{e-mail
address: {\tt mak411@psu.edu}}, Parampreet Singh$^1$\footnote{e-mail
address: {\tt singh@gravity.psu.edu}} and Aureliano
Skirzewski$^2$\footnote{e-mail address: {\tt skirz@aei.mpg.de}}
\\
\vspace{0.5em}
$^1$Institute for Gravitational Physics and Geometry,
The Pennsylvania State
University,\\
104 Davey Lab, University Park, PA 16802, USA\\
\vspace{0.5em}
$^2$Max-Planck-Institut f\"ur Gravitationsphysik, Albert-Einstein-Institut,\\
Am M\"uhlenberg 1, D-14476 Potsdam, Germany
\vspace{1.5em}
\end{center}
}

\setcounter{footnote}{0}

\newcommand{\case}[2]{{\textstyle \frac{#1}{#2}}}
\newcommand{\lP}{l_{\mathrm P}}
\newcommand{\be}{\begin{equation}}
\newcommand{\ee}{\end{equation}}
\newcommand{\bq}{\begin{eqnarray}}
\newcommand{\eq}{\end{eqnarray}}

\newcommand{\md}{{\mathrm{d}}}

\newcommand*{\R}{{\mathbb R}}
\newcommand*{\N}{{\mathbb N}}
\newcommand*{\Z}{{\mathbb Z}}
\newcommand*{\Q}{{\mathbb Q}}
\newcommand*{\C}{{\mathbb C}}
\def\f{\frac}
\def\t{\tilde}
\def\H{{\cal H}}

\begin{abstract}
 Cosmological perturbation equations are derived systematically in a
 canonical scheme based on Ashtekar variables. A comparison with the covariant
 derivation and various subtleties in the calculation and choice
 of gauges are pointed out. Nevertheless, the treatment is more
 systematic when correction terms of canonical quantum gravity are to
 be included. This is done throughout the paper for one example of characteristic
 modifications expected from loop quantum gravity.
\end{abstract}

\section{Introduction}

The backbone of most of current cosmology is the theory of
perturbation equations for metric modes around an isotropic
space-time \cite{CosmoPert}. It is used in particular for
cosmological structure formation and for testing alternative
theories beyond general relativity such as quantum gravity
candidates. The underlying equations of typical interest are the
linearized Einstein's equations, and so it is straightforward to
include corrections if they come from a Lagrangian modified by
quantum or other effects. This is, in fact, the situation
encountered in most studies of so-called trans-Planckian issues
for the effect of quantum gravity on structure formation.
Modifications derived from a Hamiltonian formulation as it is used
in canonical quantizations can, however, not be implemented in
this direct way. Since several effective modifications to
Hamiltonians have been derived in recent years in particular
within the framework of loop quantum gravity
\cite{Rov,ThomasRev,ALRev,LivRev}, it is of interest to re-derive
cosmological perturbation equations in a purely Hamiltonian
fashion starting from the gravitational Hamiltonian. This is done
in detail in this paper in a derivation based, as loop quantum
gravity itself, on real Ashtekar variables
\cite{AshVar,AshVarReell}. We will present a detailed derivation
for scalar modes in longitudinal gauge around a spatially flat
model, pointing out several subtleties compared to the Lagrangian
derivation. Our analysis treats gravitational and matter terms on
the same footing, showing how all of them can be obtained from the
total Hamiltonian. This presents the classical basis for a
systematic investigation of effective perturbation equations and
cosmological implications resulting from canonical quantum
gravity. In our calculations, only one type of corrections (from
inverse powers of metric components in the Hamiltonian) is used,
and their implications are discussed.

\section{Variables and Equations}

To set up linear metric perturbations \cite{CosmoPert}, one perturbs
the background metric
\begin{equation}
 \md s^2= a^2(\eta) \left(-\md\eta^2+\delta_{ab} \md
x^a \md x^b \right)\,,
\end{equation}
here chosen as a flat isotropic metric written in conformal time
$\eta$ and with spatial coordinates $x^a$. There are initially ten
perturbation functions for the ten metric components, but some of
them can be absorbed simply by redefining coordinates. The
remaining functions, in gauge-invariant combinations, comprise
scalar, vector and tensor modes.  We are here primarily interested
in scalar modes which in longitudinal gauge lead to a perturbed
metric
\begin{equation}\label{MetricPert}
 \md s^2= a^2(\eta) \left( -(1+2\phi) \md\eta^2
+(1-2\psi) \delta_{ab} \md x^a \md x^b \right)
\end{equation}
which is thus diagonal. Moreover, in the absence of anisotropic stress
it is consistent with longitudinal gauge to set $\phi=\psi$, reducing
the perturbations to a single function. We will also do so in our
final equations, but not immediately since it is a consequence of
equations of motion and should not be used in the process of deriving
such equations.  Expanding Einstein's equations to linear order in
$\psi$ then leads to
\begin{eqnarray}
  \nabla^2\psi-3\frac{\dot{a}}{a}\dot{\psi}-3\frac{\dot{a}^2}{a^2}\psi
 &=& - \, \frac{\kappa}{2}a^2 \delta T^0_0 \label{Pert1}\\
\ddot{\psi}+3\frac{\dot{a}}{a}\dot{\psi}+
2\left(\frac{\dot{a}}{a}\right)^{.}\psi+
\frac{\dot{a}^2}{a^2}\psi &=& \frac{\kappa}{2}a^2\delta T^a_a \label{Pert2}\\
\partial_a(\dot{\psi}+\frac{\dot{a}}{a}\psi) &=& -\,
\frac{\kappa}{2}a^2\delta T^0_a \label{Pert3}
\end{eqnarray}
where $\kappa=8\pi G$ is the gravitational constant and a dot denotes
a derivative by conformal time $\eta$. The source terms on the right
hand side of these equations are components of the energy momentum
tensor provided by the matter ingredients, also perturbed
linearly. These components follow from functional derivatives of the
matter Lagrangian by metric components, for a scalar field $\varphi$
with potential $V(\varphi)$ and Lagrangian
\[
 L_{\varphi}=-\int\md^3x \sqrt{-\det
g}\left(\frac{1}{2}g^{ab}\partial_a\varphi\partial_b\varphi+
V(\varphi)\right)
\]
we have, for instance,\footnote{Note that the
sign of the relation between pressure and $T^a_a$ depends on the
signature of the metric. Our signs correspond to positive spatial
metric components as in (\ref{MetricPert}).}
\begin{eqnarray}
 \delta T^0_0 &=& -\f{1}{a^2} \left(\dot{\bar{\varphi}} \,
\delta \dot \varphi  - \dot{\bar{\varphi}}^2 \, \psi + a^2
V,_{\varphi}(\varphi) \, \delta
\varphi \right)\\
 \delta T^0_a &=& -\f{1}{a^2} \, \dot{\bar{\varphi}} \,
\delta \varphi,_{a} \label{Tcomp}\\
 \delta T^a_b &=& \f{1}{a^2} \left(- \dot{\bar{\varphi}} \,
\delta \dot \varphi
+ \dot{\bar{\varphi}}^2 \, \psi \, + \, a^2 \,
V,_{\varphi}(\varphi) \, \delta \varphi \right)\, \delta^a_b\,.
\end{eqnarray}

In addition, the matter Lagrangian determines equations of motion for
matter fields such as those of a scalar $\varphi$:
\begin{equation} \label{phiback}
 \ddot{\bar \varphi} + 2 \frac{\dot{a}}{a} \dot{\bar{\varphi}} + a^2 \, \,
V_{,\varphi}(\bar\varphi) = 0
\end{equation}
is the background Klein-Gordon equation, whereas
\begin{equation} \label{phipert}
\delta \ddot \varphi + 2 \frac{\dot{a}}{a} \delta \dot \varphi - \nabla^2
\delta \varphi + a^2 V,_{\varphi\varphi}(\bar \varphi)\delta \varphi
+2a^2 V,_{\varphi}(\bar \varphi) \psi
-4\dot{\bar\varphi}\dot\psi=0
\end{equation}
describes the perturbed part of the scalar field.

\subsection{Canonical formalism}

Cosmological perturbation equations are the Einstein's equations
expanded in metric perturbations. Once a gauge is chosen and modes
of interest are selected, the perturbed metric is specified and
ready to be inserted in the expansion. Since the canonical
formalism is equivalent to the Lagrangian one which yields
Einstein's equations as the Euler--Lagrange equations of the
Einstein--Hilbert action, the same perturbation equations must
result. However, some of the derivations are more subtle since one
has to fix gauges and select modes at the right places. Moreover,
one first starts with a different set of variables and first order
differential equations, which are combined to the usual second
order equations. Keeping in mind that quantum gravity can lead to
several modifications it is helpful to go through the canonical
derivation in detailed steps, which is what we do in this paper.

In a canonical formulation \cite{ADM}, the Hamiltonian $H$ rather than
Lagrangian $L$ is the basic dynamical object, determining equations of
motion of any phase space function $f$ by means of Poisson brackets,
$\dot{f}=\{f,H\}$. The Poisson structure defines the kinematical
arena, which is usually written in terms of a set of basic canonical
variables such as position and momentum in mechanics. While dynamics
as well as expressions for momenta follow from the same object in a
Lagrangian formulation, they are separate in a Hamiltonian one. The
Poisson structure is thus prescribed independently of the Hamiltonian,
but both of them are needed to determine dynamics. Basic configuration
variables in a Lagrangian formulation of gravity are the components of
the space-time metric $g_{ab}$, and their momenta are determined as
usually by derivatives $\pi^{ab}=\delta L/\delta \dot{g}_{ab}$. The
dot refers here to a time coordinate in which the action is
written. Since general relativity is covariant under arbitrary changes
of space-time coordinates, the choice of time does not play a physical
role. Nevertheless, by definition of its kinematical objects a
canonical formulation does not appear manifestly covariant. Indeed,
not all components of the space-time metric appear equal: some of
them, the time-time component $N$ and the time-space components $N^a$
do not occur as first order derivatives in the action such that their
momenta vanish identically, $\pi_N=0=\pi_{N^a}$.

This is a consequence of general covariance and implies the existence
of constraints. Since momenta of $N$ and $N^a$ vanish, their equations
of motion imply
\begin{eqnarray*}
 0 &=& \dot{\pi}_N=\{\pi_N,H\}=-\delta H/\delta N \quad\mbox{ and}\\
 0 &=& \dot{\pi}_{N^a}=\{\pi_{N^a},H\}=-\delta H/\delta N^a
\end{eqnarray*}
as constraint equations on the remaining phase space variables. In
fact, because there is no absolute meaning to the time coordinate at
all, the total Hamiltonian is a sum of constraints $H=H[N]+D[N^a]$
with the Hamiltonian constraint $H[N]=\int\md^3x N(x)\delta H/\delta
N$ and the diffeomorphism constraint $D[N^a]=\int\md^3x N^a(x) \delta
H/\delta N^a$. Coordinate time evolution through Hamiltonian equations
of motion is completely specified only when $N$ and $N^a$ are known as
functions on space-time. However, there are no equations of motion for
$N$ and $N^a$ themselves; they are not dynamical since their momenta
vanish. They have to be chosen in order to fix the gauge in which
space-time properties are computed. That the constraints generate
coordinate changes can more easily be seen for the diffeomorphism
constraint which satisfies $\{f,D[N^a]\}={\cal L}_{N^a}f$ for any
phase space function $f$, where on the right hand side the Lie
derivative occurs. Both constraints receive contributions from
gravitational fields (the spatial metric and their momenta) and matter
fields.

The Hamiltonian formulation is thus based on phase space coordinates
given by the spatial metric components, matter fields and their
momenta. In addition, there are the lapse function $N$ and shift
vector $N^a$ which need to be chosen for a particular gauge. Their
dynamical behavior is given by Hamiltonian equations of motion derived
through Poisson brackets with the constraints. Since the Hamiltonian
is usually, and in particular in gravity, a quadratic polynomial of the
momenta conjugate to metric components, Poisson brackets between
configuration variables and the Hamiltonian are linear in
momenta. Thus, the Hamiltonian equations of motion for configuration
variables relate momenta to first order time derivatives. Equations of
motion for the momenta can then be reformulated as second order
differential equations for configuration variables which agree with
the Euler--Lagrange equations. Moreover, one can replace momenta in
the constraint equations by first order derivatives of configuration
variables, giving additional first order differential equations. The
set of equations (\ref{Pert1}), (\ref{Pert2}) and (\ref{Pert3}) thus
consists, from the Hamiltonian perspective, of two constraint
equations, the Hamiltonian constraint (\ref{Pert1}) and the
diffeomorphism constraint (\ref{Pert3}) and one equation of motion
(\ref{Pert2}) for the single scalar mode.

However, this set of equations is not the most general one for a
linearized metric. Gauge choices and a selection of modes have
been made, the latter excluding vector and tensor modes and
equating the lapse perturbation with the scalar mode. These put
conditions on the variables and on the multipliers $N$ and $N^a$.
In a Hamiltonian formulation one has to be careful about when to
make such choices in the process of deriving the equations of
motion. Gauge choices have to be made from the start because this
determines what the time variable and other coordinates in the
resulting differential equations mean. For instance, the
homogeneous mode of the lapse function has to be specified, which
is usually chosen as $N=1$ for proper time or $N=a$ for conformal
time. But this is to be done only for equations of motion, not for
a derivation of the constraint. It is clear that setting $N=1$ or
equal to the scale factor does not result in the right Hamiltonian
constraint $\delta H/\delta N$ which requires $N$ to be an
independent variable. Similarly, one often sets the shift vector
to zero, while the diffeomorphism constraint $\delta H/\delta N^a$
is to be imposed fully.

The correct procedure is as follows: To derive constraint equations,
no gauge choices are to be made. In fact, without knowing the
constraints it is impossible to know what gauge freedom one has. In
the next step, one derives Hamiltonian equations of motion for the
phase space variables, $q_{ab}$ and their momenta in the case of
gravity. Here, the gauge has to be chosen before computing Poisson
brackets to give meaning to coordinate derivatives. One can also
restrict to specific modes, but other conditions are not to be
done. For instance, equating the lapse perturbation to the scalar mode
is only justified as the result of equations of motion. Doing this
before computing Poisson brackets would introduce erroneous relations
between independent degrees of freedom. Thus, such a simplification
must be made only in the final expressions for equations of motion.

\subsection{Perturbed canonical variables}

Also the set of canonical variables matters for a quantization: while
classically one can change variables by canonical transformations,
their quantum representations can appear very different. Loop quantum
gravity crucially depends on properties of Ashtekar variables
\cite{AshVar,AshVarReell} due to their transformation
properties. First, one introduces a co-triad $e_a^i$ instead of the
spatial metric $q_{ab}$, related to it by $e_a^ie_b^i=q_{ab}$. (Unlike
the position of spatial indices $a,b,\ldots$, the upper or lower
positions of indices $i$ are not relevant, and summing over $i$ is
understood even though it appears twice in the same position.) An
oriented co-triad contains the same information as a metric but has
more components as it is not a symmetric tensor. This corresponds to
freedom one has in rotating the triple of triad co-vectors which does
not change the metric. Not being of geometrical relevance, this
freedom is removed in a canonical formalism by implementing the Gauss
constraint introduced below. By inverting the matrix $(e_a^i)$, one
obtains the triad $e^a_i$, a set of vector fields related to the
inverse metric by $e^a_ie^b_i=q^{ab}$. Just as the metric determines a
compatible Christoffel connection $\Gamma_{ab}^c$, a triad determines
a compatible spin connection
\begin{equation} \label{SpinConnFull}
 \Gamma_a^i= -\epsilon^{ijk}e^b_j (\partial_{[a}e_{b]}^k+
 {\textstyle\frac{1}{2}} e_k^ce_a^l\partial_{[c}e_{b]}^l)\,.
\end{equation}
Its components define the Ashtekar connection $A_a^i=\Gamma_a^i+\gamma
K_a^i$ together with those of extrinsic curvature
\begin{equation}
K_a^i=e^b_iK_{ab}=-\frac{1}{2N}e^b_i({\cal L}_tq_{ab}+D_{(a}N_{b)})
\end{equation}
where the right hand side uses lapse function $N$ and shift vector $N^a$ in
addition to the spatial metric $q_{ab}$. Moreover, ${\cal L}_t$
denotes a Lie derivative along a timelike vector field chosen to
describe changes in coordinate time, and $D_a$ is the covariant
derivative compatible with the spatial metric $q_{ab}$. In $A_a^i$, we
have the positive real Barbero--Immirzi parameter $\gamma$
\cite{AshVarReell,Immirzi} which we keep here for generality although
it will not play a large role later on.

The Ashtekar connection is thus a measure for curvature, and spatial
metric information is described by the densitized triad
$E^a_i=\left|\det e_b^j\right|e^a_i$ obtained from the triad. By
multiplying the triad by the determinant of the co-triad, which is
identical to the square root of the determinant of the spatial metric,
it becomes canonically conjugate to the Ashtekar connection:
\begin{equation}
 \{A_a^i(x),E^b_j(y)\} = \gamma\kappa \delta^b_a\delta^i_j\delta(x,y)\,.
\end{equation}
This follows from the gravitational action which in a first order
formulation contains time derivatives of the connection in the term
\[
 \frac{1}{\gamma\kappa}\int\md^3x \frac{\md A_a^i}{\md t}E^a_i
\]
showing that the connection and the densitized triad are canonically
conjugate.

An unperturbed isotropic triad and connection for flat spatial slices
\cite{IsoCosmo} can always be chosen of diagonal form
$E^a_i=p\delta^a_i$ and $A_a^i=c\delta_a^i$ with canonically conjugate
$c$ and $p$: $\{c,p\}=\frac{1}{3}\gamma \kappa$. In more familiar
terms, these variables are related to the scale factor $a$ by
$|p|=a^2$ and $c=\gamma{\md a/\md t}$. With inhomogeneous perturbations, the
triad and connection components will become space-dependent, but not
in a way that is completely unrelated between $E^a_i$ and $A^i_a$
because of the spin connection. This implies that not both the triad
and the connection can remain diagonal even when only scalar
perturbations in longitudinal gauge are considered for which the
spatial metric is diagonal. (This happens generally in inhomogeneous
situations; see also \cite{SphSymm}.) Another way to see this is by
looking at the Gauss constraint
\begin{equation} \label{Gauss}
\partial_aE^a_i+\epsilon^{ijk} E^a_jA_a^k=0
\end{equation}
which ensures invariance of physical results under rotations of the
triad, which do not change the metric. Were both $E^a_i$ and $A_a^i$
diagonal, the second term $\epsilon^{ijk} E^a_jA_a^k$ would vanish,
constraining inhomogeneity by $\partial_aE^a_i=0$.

It is most useful to keep the triad diagonal since this simplifies the
classical calculations, and even more so the quantum ones where
currently only situations of diagonal triads are sufficiently
accessible by explicit calculations. We thus introduce the perturbed
triad\footnote{In \cite{InhomLattice,QuantCorrPert} the triad
coefficient has been called $\tilde{p}(x)$ and distinguished from a
rescaled $p(x)$. The difference is not relevant for most purposes here
since our equations are mainly rescaling-invariant. Only for a
discussion of scale-dependence of correction functions is it necessary
to distinguish between these variables, as we will discuss later. We
thus drop the tilde in order not to overload the notation.}
\begin{equation} \label{TriadPert}
E^a_i=p(x)\delta^a_i=(\bar{p}+\delta p(x))\delta_i^a
\end{equation}
which gives rise to a spatial metric of the form
$q_{ab}=|\bar{p}+\delta p|\delta_{ab}$. Here and elsewhere, we split
off the background part
\begin{equation} \label{barp}
 \bar{p}:=\frac{1}{V_0}\int p(x)\md^3x
\end{equation}
with the spatial coordinate volume $V_0=\int\md^3x$. The latter is
assumed to be finite, using a compact torus topology of space, but
will not appear in the final classical equations. Using $\bar{p}$, we
then define the perturbation
\begin{equation} \label{deltap}
 \delta p(x):=p(x)-\bar{p} \quad\mbox{ such that } \int\delta
p(x)\md^3x =0\,.
\end{equation}

In the usual notation using the scale factor and the scalar metric
mode $\psi$ in longitudinal gauge we have
$q_{ab}=a^2(1-2\psi)\delta_{ab}$, leading to the identification
$|\bar{p}|=a^2$ and $\delta p=-2\bar{p}\psi$. Remaining space-time
metric components are the lapse function $N=\bar{N}+\delta N$ and the
shift vector $N^a$. Their usual notation $N=a(1+\phi)$ gives
$\bar{N}=a$, $\delta N=a\phi$ for the lapse function in conformal time
gauge, while $N^a$ is zero for scalar modes in the longitudinal
gauge. In the absence of anisotropic stress, i.e.\ non-diagonal
components of the spatial part of the energy-momentum tensor, $\phi$
is not independent of $\psi$ but has to agree with it. We will make
this identification in the final equations, but have to keep $N$ as
well as $N^a$ as free Lagrange multipliers in initial steps.

With a diagonal triad, the connection cannot be diagonal in
inhomogeneous situations. In fact, for a perturbed triad
(\ref{TriadPert}) one can compute the spin connection to be
\begin{equation} \label{SpinConn}
 \Gamma_a^i= \frac{1}{2}\epsilon_a{}^{ij} \frac{\partial_j\delta
p}{\bar{p}+\delta p}
\end{equation}
which is antisymmetric and thus non-diagonal. The diagonal part of
$A_a^i$ is then contributed solely by extrinsic curvature, which,
again in longitudinal gauge where the shift vector
vanishes,\footnote{As a gauge choice, this is to be used with care in
deriving equations of motion, as described before. In this case, one
can check that equations of motion of non-diagonal triad terms only
express non-diagonal extrinsic curvature terms through
$D_{(a}N_{b)}$. Thus, the shift vector can be consistently set to
zero.}  is proportional to a time derivative of the triad and thus
diagonal,
\be%
\label{ExtrCurv} K_a^i=k(x)\delta_a^i=(\bar k +\delta k(x)) \delta_a^i.
\ee%
 A perturbed connection then has the form
\begin{equation} \label{ConnPert}
 A_a^i=\gamma(\bar{k}+\delta k(x))\delta_a^i+ \Gamma_a^i(x)
\end{equation}
split into the perturbed but diagonal extrinsic curvature part and the
non-diagonal (in fact, antisymmetric) part coming from the perturbed
spin connection.  The direct calculation (\ref{SpinConn}) can easily
be seen to solve the Gauss constraint identically. The gravitational
variables $\bar{p}$, $\bar{k}$, $\delta p$ and $\delta k$ are thus
constrained only by the diffeomorphism and Hamiltonian constraints.

\subsection{Constraints}

The gravitational contributions to these constraints in terms of
Ashtekar variables are the diffeomorphism constraint
\begin{equation} \label{DiffConstr}
 D_G[N^a]= \frac{1}{\gamma\kappa}\int\md^3x N^aF_{ab}^iE^b_i
\end{equation}
where $F_{ab}^i= \partial_a A_b^i- \partial_b A_a^i+\epsilon^{ijk}
A_a^j A_b^k$ is the curvature of the Ashtekar connection, and the
Hamiltonian constraint
\begin{equation}\label{HamConstr}
 H_G[N] = \frac{1}{2\kappa} \int_{\Sigma} \mathrm{d}^3x N\left|\det
   E\right|^{-1/2} \left(\epsilon_{ijk}F_{ab}^iE^a_jE^b_k
  -2(1+\gamma^{2})  K_a^i K_b^j E^{[a}_iE^{b]}_j\right)\,.
\end{equation}
Note that, strictly speaking, (\ref{DiffConstr}) is the so-called
vector constraint, i.e.\ a combination of the diffeomorphism
constraint which generates spatial diffeomorphisms and the Gauss
constraint (\ref{Gauss}). Since by our choice of variables we are
solving the Gauss constraint identically, we can use
(\ref{DiffConstr}) as the diffeomorphism constraint.

\begin{figure}
\centerline{\includegraphics[width=9cm,angle=270,keepaspectratio]{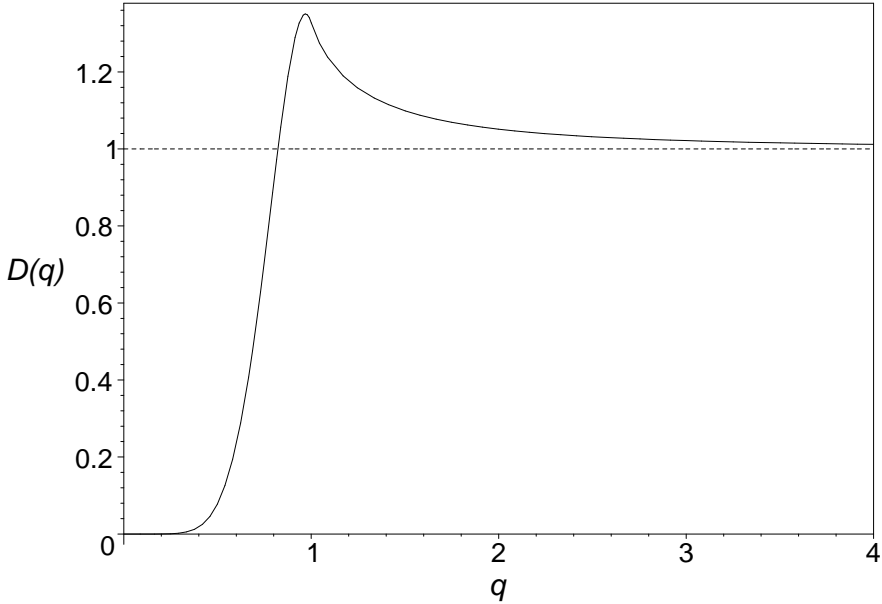}}
\caption{Typical behavior of correction functions $\alpha$,
$\beta$, $D$ and $\sigma$ which approach one from above for large
arguments. For small arguments, the functions are increasing and
reach a peak value larger than one. \label{Fig:D}}
\end{figure}

Clearly, the matrix symmetry of (\ref{TriadPert}), (\ref{SpinConn})
and (\ref{ExtrCurv}) will lead to simplifications of the Hamiltonian
and other constraints. Since constraints are also the places where
quantum corrections enter, in particular in (\ref{HamConstr}) which
determines dynamics, one should use simplifications only after such
corrections have been implemented. Specifically, in the effective
regime, the terms containing inverse powers of the scale factor,
e.g. the spin connection (\ref{SpinConn}) or $|\det E|^{-1/2}$, will
have quantum corrections. If one uses (\ref{TriadPert}) first, then
the inverse determinant in (\ref{HamConstr}) would {\em cancel} at the
classical level already, thus not acquiring any effective corrections
upon quantization. Reliable places for correction terms can thus only
be found in the full expressions. In this paper, we will accordingly
insert a function of triad components $\alpha$ to take into account
possible corrections coming from $|\det E|^{-1/2}$ and a function
$\beta$ multiplying spin connection components (\ref{SpinConn}) to
take into account the inverse triad component there.\footnote{The spin
connection is in general more complicated as obtained directly from
(\ref{SpinConnFull}). However, the spin connection is not quantized in
the full theory because it is not a covariant object and does not have
physical meaning without gauge fixing. There is thus no tight
prescription on how to obtain correction terms in perturbative
situations. A further difference between our perturbative treatment
here and the full theory is that we treat extrinsic curvature and the
spin connection separately as it has been proven useful in homogeneous
\cite{Spin} and midi-superspace models \cite{SphSymm,SphSymmHam}. This
is not done in the full theory where one rather quantizes extrinsic
curvature components in the constraint using $K_a^i
 \propto \left\{A_a^i,\left\{\int\md^3x \epsilon^{ijk}F_{ab}^i
 \frac{E^a_jE^b_k}{\sqrt{|\det
 E|}},\int{\sqrt{|\det E|}}\mathrm{d}^3x\right\}\right\}$
\cite{QSDI}. Such a treatment is also possible to analyze in our
situation but would be more complicated. We will later see that
corrections from the spin connection are much less important than
corrections from the inverse determinant explicit in the
Hamiltonian constraint. Thus, the precise treatment of extrinsic
curvature and the spin connection does not seem crucial for
phenomenology.}  These functions depend on triad components and
classically $\alpha=\beta=1$ in which case we will indeed
reproduce the classical equations. We will, however, keep $\alpha$
and $\beta$ in our equations without specifying them to
demonstrate how quantum corrections can easily propagate in more
complicated terms when equations of motion are derived; such
functions can be found in \cite{QuantCorrPert}. The typical
behavior is similar to the function $|\bar{p}|^{3/2}d(\bar{p})$
\cite{Ambig,ICGC} used in isotropic models and sketched in
Fig.~\ref{Fig:D}. For small arguments these functions are
increasing, starting from a value smaller than one which is zero
in homogeneous models and for perturbative inhomogenities but can
be non-zero if non-Abelian features of the full theory and
coherent state effects are considered \cite{BoundCoh,DegFull}. At
an intermediate scale, whose value depends on details of the
quantization, they peak at a height larger than one and then
approach the classical value one from above. The small-$\bar{p}$
behavior is thus strongly modified by non-perturbative effects
while the large-$\bar{p}$ behavior is perturbative of the form
\begin{equation} \label{alphapert}
 \alpha(\bar{p})=1+c\left(\frac{\ell_{\rm P}^2}{\bar{p}}\right)^n
\qquad \mbox{ with } c>0 \mbox{ and } n>0
\end{equation}
where $\ell_{\rm P}=\sqrt{\hbar G}$ is the Planck length.  The values
of $c$ and $n$ also depend on the quantization but are generally
positive. This allows the discussion of characteristic qualitative
effects. It is safest to use perturbation theory above the peaks of
correction functions.

The constraint (\ref{HamConstr}) can be simplified, if expressed
in terms of the spin connection and extrinsic curvature, using
$A_a^i=\Gamma_a^i+\gamma K_a^i$. The first term is then written as
\bq \label{Curv}%
F_{ab}^i &=& 2\partial_{[a}\Gamma_{b]}^i +2 \gamma\partial_{[a}
K_{b]}^i+\epsilon_{ijk}\left(\Gamma_a^j+\gamma
K_a^j\right)\left(\Gamma_b^k+\gamma K_b^k\right) \nonumber \\
&=&2\partial_{[a}\Gamma_{b]}^i +2 \gamma\partial_{[a} K_{b]}^i+\gamma
\epsilon_{ijk} \left(\Gamma_a^j K_b^k+\Gamma_b^k K_a^j
\right)+\epsilon_{ijk}\left(\Gamma_a^j\Gamma_b^k+\gamma^2 K_a^j
K_b^k\right)\,.
\eq%
Contracting with the triad, we see that the second term
\be%
\epsilon_{ijk}\partial_a K_b^i E_j^a E_k^b \propto \epsilon_{ijk}
\delta_b^i \delta_j^a \delta_k^b \propto
\epsilon_{ijk}\delta_k^i=0,\nonumber
\ee%
whereas the cross-term
\[
\epsilon_{ijk}\epsilon_{ilm}\left(\Gamma_a^l K_b^m+\Gamma_b^m
K_a^l \right) E_j^a E_k^b =2 \delta_{[l}^j \delta_{m]}^k E_j^a
E_k^b \Gamma_a^l K_b^m
\propto\Gamma_j^j K_k^k - \Gamma_j^k K_k^j
\]
vanishes because $K$ is diagonal and $\Gamma$ is anti-symmetric.
Also, the last curvature term in (\ref{Curv}) (quadratic in $K$), can
be combined with the last term of (\ref{HamConstr}) yielding
\begin{equation}\label{HamShort}
 H_G[N] = \frac{1}{\kappa} \int_{\Sigma} \mathrm{d}^3x N \alpha
\left|\det
   E\right|^{-1/2} \left(\epsilon_{ijk} \partial_a\left(\beta\Gamma_b^i\right)+
  \beta^2\Gamma_a^j \Gamma_b^k-K_a^j K_b^k \right)E^{[a}_jE^{b]}_k\,.
\end{equation}
As already announced, here we have included quantum correction
functions $\alpha$ and $\beta$ which depend on the densitized triad
and reproduce classical behavior for $\alpha=\beta=1$.

In the diffeomorphism constraint we have similar simplifications and
some terms do not contribute after writing out the curvature
components explicitly. In our variables,
\begin{eqnarray*}
 F_{ab}^i E_i^b&=& 2\gamma(\bar{p}+\delta p)\partial_a(\bar{k}+\delta
k)+ \epsilon_{bjk}(\Gamma_a^j+\gamma K_a^j)\Gamma_b^k (\bar{p}+\delta
p)\\
 &=& 2\gamma(\bar{p}+\delta p)\partial_a(\bar{k}+\delta
k)- (\Gamma_a^j+\gamma K_a^j)\partial_j\delta p
\end{eqnarray*}
using (\ref{SpinConn}) in the last line. Then, with
$\Gamma_a^j\partial_j\delta p=0$ the constraint
\[
 D[N^a]= \frac{1}{\kappa}\int\md^3x \delta N^a(2\bar{p}
\partial_a\delta k-\tilde{\beta} \bar{k}\partial_a\delta
p)
\]
for $\bar{N}^a=0$ results. Note that we did use the expression for the
spin connection in the second term which could thus receive a correction
function $\tilde{\beta}$. Since it is coming from the spin connection
it should equal $\beta$, but we keep it separate because the
diffeomorphism constraint is not quantized in infinitesimal form in
loop quantum gravity. One rather implements finite diffeomorphisms
exactly \cite{ALMMT} such that no corrections are expected. As we will
see later, effects of $\tilde{\beta}$ are not important such that the
precise prescription here does not matter much.  Moreover, there is no
inverse determinant in the diffeomorphism constraint and no need for a
correction function $\alpha$.

 Together with contributions from matter fields, this defines
constraints on the basic variables. For a scalar field $\varphi$ with
momentum $\pi$ and potential $V(\varphi)$, we have a contribution
\begin{equation}
 D_{\varphi}[N^a]= \int\md^3x N^a\pi\partial_a\varphi
\end{equation}
to the diffeomorphism constraint and a contribution by the matter
Hamiltonian
\begin{equation}
 H_{\varphi}[N]= \int\!\!\md^3x
N\left(\frac{1}{2}D \frac{\pi^2}{\sqrt{|\det E|}}+
\frac{1}{2}\sigma\frac{E^a_iE^b_i \partial_a\varphi
\partial_b\varphi}{\sqrt{|\det E|}}+ \sqrt{|\det E|}V(\varphi)\right)
\end{equation}
to the Hamiltonian constraint. Again, only the contribution to the
Hamiltonian constraint contains inverse powers which occur in two
different forms. The kinetic part has a single inverse determinant
which we correct by a function $D$ (as it has been used in isotropic
models), while the gradient term has additional triad components in
the numerator which can lead to a different correction function
$\sigma$. Note also that, while formally the combination of triad
components in the gradient term is the same as that in the
gravitational part of the Hamiltonian constraint, it is only the
symmetric part in $a$ and $b$ which enters the matter gradient term
but the antisymmetric part in the gravitational constraint. We
therefore keep the correction functions $\alpha$ and $\sigma$
independent.

The equations to consider are thus the
two constraint equations
\begin{eqnarray}
 D[N^a] &=& D_G[N^a]+D_{\varphi}[N^a]=0\\
 H[N] &=& H_G[N]+H_{\varphi}[N]=0
\end{eqnarray}
together with the Hamiltonian equations of motion
\begin{equation} \label{EqMot}
 \dot{f}=\{f,H[N]+D[N^a]\}
\end{equation}
for any of the variables $\bar{p}$, $\bar{k}$, $\bar{\varphi}$,
$\bar{p}_{\varphi}$ and the fields $\delta p$, $\delta k$,
$\delta\varphi$ and $\delta \pi$. The components $N$ and $N^a$, i.e.\
$\bar{N}$, $\delta N$ and $\delta N^a$ play the role of Lagrange
multipliers for the constraints. When computing the Poisson brackets
in equations of motion (\ref{EqMot}) one has to keep these multipliers
as independent variables at this stage, as discussed before. Only
$\bar{N}$ has to be specified to fix the time gauge, with the two most
common choices $\bar{N}=1$ for proper time and $\bar{N}=a$ for
conformal time which we will use here. The fields $\delta N$ and
$\delta N^a$, on the other hand, must not yet be fixed to $\psi$ or
zero, respectively, but be kept independent of the canonical fields.

\section{Linear perturbation}

In order to derive linearized equations of motion, we expand the
Hamiltonian to second order in the field perturbations so as to
get linear equations after taking Poisson brackets. For the
constraint equations themselves, the linear coefficients of
$\delta N$ and $\delta N^a$ will result as perturbation equations,
accompanied by equations of motion for $\delta p$, $\delta k$,
$\delta\varphi$ and $\delta \pi$, which can be combined to two
second order differential equations of motion for $\delta p$ and
$\delta\varphi$. Spatial integrations of terms linear in
perturbations give zero because we have split off the homogeneous
background contributions explicitly in definitions such as
(\ref{deltap}) for $\delta p$.  The Poisson structure for these
variables can then be computed by inserting them in the full
action term
\[
 \frac{1}{\gamma\kappa}\int\md^3x \frac{\md A_a^i}{\md t}E^a_i=
\frac{3}{\kappa}\int\md^3x \frac{\md k(x)}{\md t}p(x)=
\frac{3V_0}{\kappa}\frac{\md\bar{k}}{\md t}\bar{p}+ \frac{3}{\kappa}
\int\md^3x \frac{\md \delta k}{\md t}\delta p
\]
such that
\begin{equation}
 \{\bar{k},\bar{p}\} = \frac{\kappa}{3V_0}\quad,\quad \{\delta
k(x),\delta p(y)\}=\frac{\kappa}{3}\delta(x,y)\,.
\end{equation}

\subsection{Variations}

All necessary equations follow from variations of the total constraint
\begin{eqnarray}
 &&H=H_G[N]+H_{\varphi}[N]+ D_G[N^a]+D_{\varphi}[N^a]\nonumber\\
 &&=\frac{1}{2\kappa}\int\md^3xN(x)\alpha(p) \left( -6(\bar{k}+\delta k)^2
|\bar{p}+\delta p|^{1/2} + \frac{1}{2}
\frac{(\beta(p)^2+4\beta^\prime(p) p - 4 \beta(p))\partial^a
\delta p \partial_a \delta p}{|\bar{p}+\delta
p|^{3/2}}\nonumber \right. \\
&&+\left. 2\frac{\beta(p)\nabla^2 \delta p}{|\bar{p}+\delta
p|^{1/2}}\right)+\frac{1}{\kappa}\int\md^3x \delta
N^a\left(2\bar{p}\partial_a\delta k- \tilde{\beta}(p)\bar{k}\partial_a\delta
p\right)+ H_{\varphi}[N]+D_{\varphi}[N^a]
\end{eqnarray}
by the independent variables $\bar{N}$, $\delta N$, $\delta N^a$,
$\bar{p}$, $\delta p$, $\bar{k}$, $\delta k$, $\bar{\varphi}$,
$\delta\varphi$, $\bar{\pi}$ and $\delta \pi$. The background
shift vector $\bar{N}^a$ does not appear because it vanishes for
homogeneous models, as does the background diffeomorphism
constraint it would be the Lagrange multiplier of. It would have
to be considered when perturbing around an inhomogeneous
background.

We then have the following equations:
\begin{equation} \label{VarBarN}
 0=\frac{\partial H[N]}{\partial\bar{N}} = -\f{3V_0}{\kappa}\alpha \sqrt{|\bar
p|}\bar k^2 + \frac{\partial H_{\varphi}[N]}{\partial\bar{N}}
\end{equation}
gives the background Friedmann equation which is corrected by higher
order terms such as $\int\md^3x\alpha\delta k^2\sqrt{|\bar{p}|}$ in
the perturbations if higher orders are retained in the expansion. In this
variation, $\bar{N}$ is kept independent because it must be varied,
but from now on we set $\bar{N}=\sqrt{|\bar{p}|}$ to choose conformal
time, derivatives of which will be denoted by a dot (while a prime is
used for $\bar{p}$-derivatives). Then,
\begin{equation} \label{VarDeltaN}
 0=\frac{\delta H[N]}{\delta(\delta N)} = \f{\sqrt{
 |\bar p|}}{\kappa}\left( -6 \alpha k \delta k - 3\alpha k^2\left(1+2 \bar p
 \f{\alpha^\prime}{\alpha}\right) \f{\delta p}{2 \bar
p}+\f{\beta}{\bar p} \nabla^2\delta p\right)+ \frac{\delta
H_{\varphi}[N]}{\delta(\delta N)}
\end{equation}
gives the first perturbation equation equivalent to (\ref{Pert1}),
\begin{equation} \label{VarDeltaNa}
 0=\kappa\frac{\delta D[N^a]}{\delta(\delta N^a)} = 2\bar{p}\partial_a\delta
k- \tilde{\beta}\bar{k}\partial_a\delta p+ \kappa
 \frac{\delta D_{\varphi}[N^a]}{\delta(\delta N^a)}
\end{equation}
gives the third perturbation equation equivalent to (\ref{Pert3}) and
\begin{equation} \label{VarBarP}
\dot{\bar{k}}= \f{\kappa}{3V_0}\frac{\partial H}{\partial\bar{p}} =
-\alpha \bar k^2 +
\frac{\kappa}{3V_0} \frac{\partial H_{\varphi}[N]}{\partial\bar{p}}
\end{equation}
gives the background Raychaudhuri equation. In
\begin{eqnarray} \label{VarDeltaP}
 \delta\dot{k}=\f{\kappa}{3}\frac{\delta H}{\delta(\delta p)} \!\!\!&=&
 \!\!\!- \f{1}{\sqrt{|\bar p|}}\left(\alpha+\alpha^\prime\bar
 p\right) \left(\delta N \bar k^2 + 2\bar N\bar k\delta k \right)+
\f{\delta p}{|\bar
 p|^{3/2}}\bar N \bar k^2 \left(\alpha- \alpha^\prime\bar p -
\alpha^{\prime \prime} \bar p^2\right)\\
&& + O\left(\delta(N\sqrt{|p|})\right)\nonumber \\
 &&-\f{\bar N}{6\bar p^{3/2}}\left(\alpha \beta (\beta-2)-
4\beta(\alpha^\prime\bar p)\right)\nabla^2 \delta p
 +\f{\alpha \beta}{3 \sqrt{|\bar p|}}\nabla^2 \delta N
+ \frac{\kappa}{3}\frac{\delta H_{\varphi}[N]}{\delta(\delta p)}\nonumber
\end{eqnarray}
which gives the second
perturbation equation equivalent to (\ref{Pert2}), the term
$O(\delta(N\sqrt{|p|}))$ indicates that there are additional terms
proportional to $\delta(N\sqrt{|p|})$ which are not evaluated
explicitly here. They will cancel exactly in the final equations for
the modes used here, but would give non-zero contributions if the
lapse perturbation and the scalar mode are not identified or if other
gauges are used.

Finally,
\begin{equation} \label{VarBarK}
\dot{\bar{p}}=-\f{\kappa}{3V_0}\frac{\partial H}{\partial\bar{k}} =
2\alpha\bar p \bar{k}
\end{equation}
relates the connection component $\bar{k}$ to the time derivative
of $\bar{p}$ or the scale factor $a$. Together with the
perturbation equation
\begin{equation} \label{VarDeltaK}
 \delta\dot{p}=-\f{\kappa}{3}\frac{\delta H}{\delta(\delta k)} = 2
\alpha \bar p\delta k
 + 2\alpha' \bar{k} \bar p \delta p
\end{equation}
which relates the connection component $\delta k$ to the time
derivative of $\delta p$,  it can be used to eliminate the
extrinsic curvature components.

For the matter variables we obtain four additional equations,
\begin{equation} \label{VarBarPhi}
 \dot{\bar
 \pi}=-\frac{1}{V_0}\frac{\partial H}{\partial\bar{\varphi}} =
-\bar p^2 V,_{\varphi}(\bar \varphi)
\end{equation}
which gives the background Klein--Gordon equation,
\begin{equation} \label{VarDeltaPhi}
\delta\dot\pi=- \frac{\delta H}{\delta(\delta\varphi)} = -\bar p
\left(V,_{\varphi}(\bar \varphi)\delta p + V,_{\varphi\varphi}(\bar
\varphi)\delta \varphi -\sigma(\bar p)\nabla^2\delta\varphi\right)
\end{equation}
which gives the matter perturbation equation,
\begin{equation} \label{VarBarPPhi}
\dot{\bar{\varphi}}= \frac{1}{V_0}\frac{\partial H}{\partial\bar{\pi}} =
\f{D(\bar p)\bar{N}}{\bar p^{3/2}} \bar \pi
\end{equation}
which relates $\bar{\pi}$ to the time derivative of $\bar{\varphi}$
and
\begin{equation} \label{VarDeltaPPhi}
  \delta\dot{\varphi}=\frac{\delta H}{\delta(\delta \pi)} =
 \f{D(\bar p)}{\bar p}\left( \delta \pi -
 \f{\bar\pi \delta p}{\bar{p}D(\bar p)}
 (2 D(\bar p) - D^\prime(\bar p)\bar p)\right)
\end{equation}
which relates $\delta \pi$ to the time derivative of $\dot{\varphi}$.

Eqs.~(\ref{VarBarK}), (\ref{VarDeltaK}), (\ref{VarBarPPhi}) and
(\ref{VarDeltaPPhi}) will be used to eliminate momenta from the
equations, rewriting some of them as second order differential
equations.


\subsection{Metric equations}

We first turn to the more complicated equations obtained by varying
with respect to metric modes. Here, both the gravitational and the
matter part of the constraints contribute, whose variations are
discussed separately. From now on, we evaluate the variation equations
only for the case $\delta N^a=0$ (for longitudinal gauge without
vector modes) and $\delta N=-\delta p/2\sqrt{|\bar{p}|}$ (identifying
$\phi=\psi$). The latter identification implies
$\delta(N\sqrt{|p|})=0$ at the linearized level which we will use from
now on.

\subsubsection{Gravitational part}

Eq.~(\ref{VarBarK}) can be rewritten as
\be\label{Hubble}%
 \H:=\f{\dot{ \bar p}}{2 \bar p} =\alpha \bar{k}
\ee%
where $\H$ is the conformal Hubble rate. Inserting it in
Eqs.~(\ref{VarBarN}) and (\ref{VarBarP}) gives the first order
Friedmann equation
\begin{equation}\label{FriedmannBG}
 \H^2=\f{\alpha\kappa }{3V_0\sqrt{|\bar{p}|}}
\frac{\partial H_{\varphi}[N]}{\partial\bar{N}}
\end{equation}
and the second order Raychaudhuri equation
\[
 \dot{\H}=-\H^2\left(1-\f{2 \alpha^\prime \bar
 p}{\alpha}\right)+\f{\alpha \kappa}{3V_0} \f{\partial
 H_\varphi[N]|_{\bar N = \sqrt{\bar p}}}{\partial \bar p}
\] for the background metric. Here, as well as in any equation of
motion, the lapse function is fixed prior to taking the
$p$-derivative. In other words, any appearance of a time
derivative implies that a time gauge has been chosen, i.e. the
lapse function has been fixed. With this in mind, the background
Raychaudhuri equation can be written
\begin{equation} \label{RaychaudhuriBG}
 \dot{\H}=-\H^2\left(1-\f{2 \alpha^\prime \bar
 p}{\alpha}\right)+\f{\alpha \kappa}{3V_0} \left(\f{\partial
 H_\varphi[N]}{\partial \bar p}+\f{\partial \bar N}{\partial \bar p}\f{\partial
 H_\varphi[N]}{\partial \bar N}\right)|_{\bar N = \sqrt{\bar p}}
\end{equation}
Solving Eq.~(\ref{VarDeltaK}) for $\delta k$ we obtain
\begin{equation}\label{deltak}
\delta k = \frac{\delta\dot{p}}{2\alpha\bar{p}}-
\frac{\alpha'}{\alpha} \bar{k}\delta p\,,
\end{equation}
and inserting $\bar{k}$ and $\delta k$ in terms of $\dot{\bar{p}}$
and $\delta \dot p$ in Eqs.~(\ref{VarDeltaN}), (\ref{VarDeltaNa})
and (\ref{VarDeltaP}) gives the perturbation equations
\begin{eqnarray}
-\frac{\alpha\beta}{3\bar{p}}\nabla^2\delta p- {\cal H}
\frac{\delta\dot p}{\bar{p}}+ {\cal H}^2(1-\alpha'\bar{p}/\alpha)
\frac{\delta p}{2\bar{p}} &=& \frac{\kappa \alpha}{3\sqrt{|\bar{p}|}}
 \frac{\delta H_{\varphi}[N]}{\delta (\delta N)}\label{Hamp}\\
\alpha^{-1}\partial_a(-\delta\dot{p}+{\cal H}\delta
p(\tilde{\beta}+2\alpha^\prime\bar{p}/\alpha))
 &=& \kappa\frac{\delta D_{\varphi}[N^a]}{\delta (\delta N^a)} \label{Diffp}\\
 \f{1}{\alpha}\delta \ddot p +\f{1}{3}\left(\alpha \beta (\beta-1)-4\beta(\alpha^\prime\bar p)\right)\nabla^2 \delta p
 - \f{\H}{\alpha}\left( 1+2\alpha^\prime\bar p/\alpha\right)\delta \dot p  && \nonumber \\
 -\f{\dot \H \alpha^\prime}{\alpha^2}\delta p-\left(\f{\H}{\alpha}\right)^2
 \left( 2\alpha^{\prime \prime} \bar p^2 + \alpha^\prime\bar p + \alpha -4(\alpha^\prime \bar
 p)^2/\alpha\right)\delta p&=&\f{2\kappa}{3}\bar p\frac{\delta
H_{\varphi}[N]}{\delta (\delta p)} \label{Motp}
\end{eqnarray}
for the metric mode $\delta p$.

\subsubsection{Matter part and energy-momentum}

Rather than using the energy momentum tensor as source, the
primary object in a canonical analysis is the Hamiltonian combined
with the diffeomorphism constraint. The matter Hamiltonian is
directly related to energy density \footnote{Although there is an
inverse triad in this equation, it is not quantized as there is no
energy density operator in an inhomogeneous setting of loop
quantum gravity. In any case, such a re-definition would just
change the relation between energy density and energy-momentum
components but not affect the primary equations
(\ref{Hamp}-\ref{Motp}) and (\ref{KG_delta_p}).}
\begin{equation} \label{EnergyDens}
 \rho_{\varphi}=\frac{1}{\sqrt{|\det E|}} \frac{\delta H_{\varphi}[N]}{\delta
N}
\end{equation}
while contributions to the diffeomorphism constraint give the energy
flux density
\begin{equation} \label{Flux}
 V_{\varphi,a}=\frac{1}{\sqrt{|\det E|}}
\frac{\delta D_{\varphi}[N^b]}{\delta N^a}\,.
\end{equation}
This corresponds to time-time and time-space components of the
energy momentum tensor. The remaining components, in the absence
of anisotropic stress, are pressure components which we use here
only in the isotropic case. From the thermodynamical definition of
pressure as $P=-\delta E_\varphi/\delta V$ with energy $E_\varphi$
and volume $V$, pressure components can then be derived through
\footnote{In this definition, in contrast with
(\ref{RaychaudhuriBG}), the lapse function is independent of the
triad.}
\begin{equation} \label{Pressure}
 P_{\varphi}= -\frac{1}{N}\frac{\delta H_{\varphi}[N]}{\delta \sqrt{|\det E|}}
\end{equation}
from the Hamiltonian.

For the perturbative treatment, we again split these expressions into
background and perturbation parts such as $\bar{\rho}$ and $\delta
\rho$. By the chain rule, we have
\begin{eqnarray}
 \rho_{\varphi}(x) &=& |p(x)|^{-3/2}\left(\frac{\delta \bar{N}}{\delta N(x)}
\frac{\partial H_{\varphi}[N]}{\partial \bar{N}}+\int\md^3y \frac{\delta(\delta
N(y))}{\delta N(x)}\frac{\delta H_{\varphi}[N]}{\delta(\delta N(y))}
\right)\nonumber\\
&=& |p(x)|^{-3/2}\left(\frac{1}{V_0}\frac{\partial
H_{\varphi}[N]}{\partial\bar{N}}+
\frac{\delta H_{\varphi}[N]}{\delta(\delta N(x))}- \frac{1}{V_0}
\int\md^3y\frac{\delta
H_{\varphi}[N]}{\delta(\delta N(y))}\right)\label{mattermodes}\\
&=& \bar{\rho}_{\varphi}+\delta\rho_{\varphi}(x) \nonumber
\end{eqnarray}
where we used
\[
 \frac{\delta\bar{N}}{\delta N(x)}=\frac{1}{V_0} \quad\mbox{ and }\quad
\frac{\delta(\delta N(y))}{\delta N(x)}=\delta(x,y)-\frac{1}{V_0}
\]
for $\bar{N}:=V_0^{-1}\int\md^3x N(x)$ and $\delta
N(x)=N(x)-\bar{N}$. The
last term in (\ref{mattermodes}) vanishes because $\delta
H/\delta(\delta N(y))$ is linear (or in general odd) in perturbations
and thus vanishes when integrated over space. The remaining terms then
define the background energy density
\begin{equation} \label{BackgroundDens}
 \bar{\rho}_{\varphi}=\frac{1}{|\bar{p}|^{3/2}V_0}\frac{\partial
 H_{\varphi}[N]}{\partial\bar{N}} =-\bar{T}^0_0
\end{equation}
and the linear perturbation
\begin{equation}
 \delta\rho_{\varphi}(x)= -\frac{3\delta p}{2|\bar{p}|^{5/2}V_0} \frac{\partial
 H_{\varphi}[N]}{\partial\bar{N}}+
 |\bar{p}|^{-3/2}\frac{\delta H_{\varphi}[N]}{\delta(\delta N(x))}=
-\delta T^0_0(x) \,.
\end{equation}
Thus,
\begin{equation} \label{HphiNN}
- \frac{\delta H_{\varphi}[N]}{\delta(\delta N)}= |\bar{p}|^{3/2}\delta
T^0_0+ \frac{3}{2}\sqrt{|\bar{p}|}\bar{T}^0_0\delta p\,.
\end{equation}

Similarly, we obtain
\begin{equation}
 \bar{V}_{\varphi,a}=\frac{1}{\bar{N}|\bar{p}|^{3/2}}
\frac{\partial D_{\varphi}[N^a]}{\partial
\bar{N}^a}=-\bar{T}^0_a
\end{equation}
which vanishes for a homogeneous background, and with this
\begin{equation} \label{flux}
 \delta V_{\varphi,a}(x)=\frac{1}{\bar{N}|\bar{p}|^{3/2}}
\frac{\delta D_{\varphi}[N^a]}{\delta (\delta
N^a(x))}= -\delta T^0_a(x)
\end{equation}
for the flux. Finally, we have
\begin{equation} \label{BackgroundPress}
 \bar{P}_{\varphi}= -\frac{2}{3\bar{N}\sqrt{|\bar{p}|}V_0}\frac{\partial
H_{\varphi}[N]}{\partial \bar{p}}=\bar{T}^a_a
\end{equation}
and
\begin{equation}
 \delta P_{\varphi}(x)= \delta\left(-\frac{2}{3N\sqrt{|p|}}
\frac{\delta H_{\varphi}[N]}{\delta (\delta p)}\right)= \delta T^a_a(x)
\end{equation}
for pressure. Note again that the lapse is treated as an
independent function at the stage of differentiation. Then using
$\delta(N\sqrt{|p|})=0$, this gives
\begin{equation} \label{HphiNp}
 \frac{\delta H_{\varphi}[N]}{\delta(\delta p)}= -\frac{3}{2}|\bar{p}|
\delta T^a_a \,.
\end{equation}

For a scalar field with correction terms in the Hamiltonian, these
formulae yield the energy-momentum components
\begin{eqnarray}
 \bar T^0_0 &=& -\f{\dot{\bar{\varphi}}^2}{2 \bar{p} D} - V(\bar\varphi)\\
  \bar T^0_a &=& 0\\
 \bar T^a_a &=& -\frac{1}{2\bar{p}D}\dot{\bar{\varphi}}^2
\left(1-\frac{2}{3}\frac{D'\bar{p}}{D}\right)+
 V(\bar\varphi)
\end{eqnarray}
for the background and
\begin{eqnarray}
 \delta T^0_0 &=& -\frac{\delta p\dot{\bar{\varphi}}^2}{2\bar{p}^2D}
\left(1-\frac{D'\bar{p}}{D}\right)- V,_{\varphi}\delta\varphi-
\frac{\dot{\bar{\varphi}} \delta\dot{\varphi}}{\bar{p}D}\\
 \delta T^0_a &=& -\f{1}{\bar{p}D} \, \dot{\bar{\varphi}} \, \delta \varphi,_{a}\\
 \delta T^a_a &=& \nonumber-\frac{\delta p\dot{\bar{\varphi}}^2}{2\bar{p}^2D}
\left(1-\frac{7}{3}\frac{D'\bar{p}}{D}+
\frac{4}{3}\left(\frac{D'\bar{p}}{D}\right)^2-
\frac{2}{3}\frac{D''\bar{p}^2}{D}\right)
-\frac{\dot{\bar{\varphi}}\delta\dot\varphi}{\bar{p}D}\left(1-
\frac{2}{3}\frac{D'\bar{p}}{D}\right)+
V,_{\varphi}\delta\varphi\\{}
\end{eqnarray}
for perturbations.

\subsection{Matter equations}

Solving Eq.~(\ref{VarBarPPhi}) for $\bar{\pi}$ in terms of
$\dot{\bar{\varphi}}$ and inserting it into Eq.~(\ref{VarBarPhi})
yields the Klein--Gordon equation
\begin{equation}
 \ddot{\bar \varphi} + \frac{\dot{\bar{p}}}{\bar{p}}
 \dot{\bar{\varphi}}\left(1-\f{D^\prime \bar p}{D}\right)
+ \bar{p} \, D \, \, V,_{\varphi}(\bar\varphi) = 0
\end{equation}
for the background scalar field $\bar{\varphi}$.

Taking a time derivative of Eq.~(\ref{VarDeltaPPhi}), one gets
\bq%
 \delta \ddot \varphi &=& \left( \f{D}{\bar p}\right)^{\dot{}}
 \left( \delta \pi + 2 \bar{\pi} \f{\delta p}{\bar p}
\left(2 - \f{D^\prime \bar
 p}{D}\right)\right)+ \f{D}{\bar p}\left( \delta \pi - 2 \bar \pi
\f{\delta p}{\bar p} \left(1 - \f{D^\prime \bar
 p}{2 D}\right)\right)^{\dot{}} \\
 &=&\left( \f{D}{\bar p}\right)^{\dot{}}\f{\bar p \delta \dot
 \varphi}{D}
 + \f{D}{\bar p}\left( \delta \dot \pi - 2
\left(1 - \f{D^\prime \bar p}{2 D}\right)\left(\dot{\bar\pi}
\f{\delta p}{\bar p}+
 \f{\bar\varphi}{2 D} \left(\delta\dot p -2 \H \delta p\right) \right)
 \right)+ \dot{\bar\varphi}  \f{\delta p}{\bar p} \left(\f{D^\prime \bar
 p}{D}\right)^{\dot{}}\nonumber
\eq
where the previous equations (\ref{VarDeltaPhi}) and
(\ref{VarBarPPhi}) have been used. Finally, substituting $\bar
\pi$ and $\delta \dot \pi$ from (\ref{VarBarPhi}) and
(\ref{VarDeltaPhi}), we arrive at the Klein-Gordon equation for
the perturbed part of the scalar field
\bq\label{KG_delta_p}%
\delta \ddot \varphi
&+& 2 \H \delta \dot \varphi \left(1 - \f{D^\prime\bar
p}{D}\right)-D \sigma \nabla^2 \delta \varphi +
D\bar p V,_{\varphi\varphi}(\bar{\varphi})\delta \varphi
+(D -D^\prime\bar p) V,_{\varphi}(\bar \varphi) \delta p \nonumber\\
&+&2\dot{\bar\varphi}\f{\delta\dot p}{\bar
p}\left(1-\f{D^\prime\bar
p}{2D}\right)-2\dot{\bar\varphi}\H\f{\delta p}{\bar p}\left(2+\bar
p^2\f{D^{\prime\prime}}{D}-\left(\f{D^\prime \bar
p}{D}\right)^2\right)=0\,.
\eq

\subsection{Translation to metric variables}

We can now finally write our equations of motion in familiar form by
replacing derivatives of the matter Hamiltonian by energy momentum
tensor components and by introducing the scalar mode $\psi=-\delta
p/2\bar{p}$. We keep the variable $\bar{p}$ rather than expressing it
as the scale factor squared since this is the basic quantity appearing
in our corrections functions from quantum gravity.

The background equations (\ref{FriedmannBG}) and
(\ref{RaychaudhuriBG}) become
\begin{equation}
 \H^2=\f{\kappa}{3}
 \alpha\bar{p}\bar{\rho}_{\varphi}
\end{equation}
and
\begin{equation}
 \dot{\H}=-\H^2\left(1-\f{2 \alpha^\prime \bar
 p}{\alpha}\right)+\f{\kappa}{6}\alpha\bar{p}\left(
\bar \rho_{\varphi}-3 \bar{P}_{\varphi}\right)
\end{equation}
using (\ref{BackgroundDens}) and
(\ref{BackgroundPress}). Eq.~(\ref{Hamp}), resulting from
(\ref{VarDeltaN}), together with (\ref{HphiNN}) yields
\be%
\alpha \beta \nabla^2 \psi - \f{3}{\alpha} \H \dot \psi -
\f{3}{\alpha} \H^2 \psi \left( 1-\f{\alpha^\prime \bar
p}{\alpha}\right)= -\f{\kappa}{2}\bar p \delta T_0^0\,,
\ee%
Eq.~(\ref{Motp}), resulting from (\ref{VarDeltaP}), together with
(\ref{HphiNp}) yields
\bq%
\ddot \psi &+&2\psi \dot \H\left(1-\f{\alpha^\prime \bar
p}{\alpha}\right)+3\dot\psi \H\left(1 - \f{2}{3} \f{\alpha^\prime
\bar p}{\alpha}\right) \nonumber + \f{\alpha \beta}{3} \nabla^2
\psi
\left(\alpha(\beta-1)- 4 \alpha^\prime \bar p\right)\\
&+&\psi \H^2 \left(1-5 \f{\alpha^\prime \bar p}{\alpha} +
4\left(\f{\alpha^\prime \bar p}{\alpha}\right)^2 - 2 \f{
\alpha^{\prime \prime} \bar p^2}{\alpha}\right) =
\f{\alpha\kappa}{2} \bar p \delta T_a^a
\eq%
and from (\ref{Diffp}) together with (\ref{flux}) we obtain
\begin{equation} \label{PertMod3}
 \partial_a\left(\dot{\psi}+{\cal
 H}\psi(2-\tilde{\beta}-2\alpha'\bar{p}/\alpha)\right)=
- \frac{\kappa}{2}
\bar{p} \delta T^0_a\,.
\end{equation}
Here, the energy-momentum tensor components are
\begin{eqnarray}
 \delta T^0_0 &=& -\frac{\dot{\bar{\varphi}}^2\psi}{2\bar{p}D}
\left(1-\frac{D'\bar{p}}{D}\right)- V,_{\varphi}\delta\varphi-
\frac{\dot{\bar{\varphi}} \delta\dot{\varphi}}{\bar{p}D}\\
 \delta T^0_a &=& -\f{1}{\bar{p}D} \, \dot{\bar{\varphi}} \, \delta \varphi,_{a}\\
 \delta T^a_a &=& \nonumber-\frac{\dot{\bar{\varphi}}^2\psi}{2\bar{p}D}
\left(1-\frac{7}{3}\frac{D'\bar{p}}{D}+
\frac{4}{3}\left(\frac{D'\bar{p}}{D}\right)^2-
\frac{2}{3}\frac{D''\bar{p}^2}{D}\right)
-\frac{\dot{\bar{\varphi}}\delta\dot\varphi}{\bar{p}D}\left(1-
\frac{2}{3}\frac{D'\bar{p}}{D}\right)+
V,_{\varphi}\delta\varphi
\end{eqnarray}
and $\varphi$ is subject to the Klein-Gordon equation
\begin{equation}
\ddot{\bar \varphi} + 2\H
 \dot{\bar{\varphi}}\left(1-\f{D^\prime \bar p}{D}\right)
+ \bar{p} \, D \, \, V,_{\varphi}(\bar\varphi) = 0\end{equation}
for the background and
\bq\label{KG_delta_psi}%
\delta \ddot \varphi &+& 2 \H
\delta \dot \varphi \left(1 - \f{D^\prime\bar p}{D}\right)-D
\sigma \nabla^2 \delta \varphi +
D\bar p V,_{\varphi\varphi}(\bar{\varphi})\delta \varphi \nonumber+
2(D -D^\prime\bar p)\bar p V,_{\varphi}(\bar \varphi) \psi \\
&-&4\dot{\bar\varphi}\dot\psi\left(1-\f{D^\prime\bar
p}{2D}\right)+4\dot{\bar\varphi}\psi\H\left(
\f{D^\prime\bar p}{D}+\f{D^{\prime\prime}\bar p^2}{D}-\left(\f{D^\prime
\bar p}{D}\right)^2\right)=0
\eq
for the perturbation, obtained by expressing (\ref{KG_delta_p}) in
terms of $\psi$.

Comparison with Eqs.\ (\ref{Pert1}), (\ref{Pert2}), (\ref{Pert3}),
(\ref{Tcomp}), (\ref{phiback}) and (\ref{phipert}) shows that the
classical equations are indeed reproduced when all correction
functions equal one. As the classical perturbation equations, the
corrected ones are scale invariant when the background is flat. This
is manifest in the written form since $\psi$ is scale invariant and,
although the background scale factor $a^2=|\bar{p}|$ appears
explicitly, any combination such as $\alpha'\bar{p}$ where the prime
denotes a derivative by $\bar{p}$ is scale invariant, too. Although
the correction functions depend on $\bar{p}$, the derivation shows
that they do so only in combinations which are scale invariant, taking
into account normalizations provided by a quantum state
\cite{InhomLattice,QuantCorrPert}.

\section{Covariance}

We have derived corrected perturbation equations in a fixed gauge,
which simplified quantum and classical calculations. Their space-time
covariance is thus not obvious, just as the classical equations in the
form (\ref{Pert1}), (\ref{Pert2}) and (\ref{Pert3}) are not
manifestly covariant. When classical equations are modified by
quantum corrections, in particular in a canonical scheme, it is not
clear whether covariance will be broken. Canonically, space-time
covariance is realized if the lapse function $N$ and shift vector
$N^a$ are not restricted by equations of motion but can be specified
freely as a gauge choice. This is always the case if the Hamiltonian
and diffeomorphism constraints form a first class set, i.e.\ their
Poisson brackets among each other vanish when the constraints are
satisfied. With arbitrary modifications in their terms, this is
unlikely to remain true, suggesting a breakdown of general covariance.

The situation in quantum gravity is, however, more general because new
quantum degrees of freedom arise, which can absorb some of the
restrictions which would otherwise be imposed on $N$ and $N^a$. In
fact, for an effective description of a canonical quantum theory
\cite{Karpacz} one derives effective constraints such as
$H_{\rm eff}=\langle\hat{H}\rangle$ and $D_{\rm
eff}=\langle\hat{D}\rangle$ as expectation values in suitable
states. Since states are described by many, in fact infinitely many,
more variables (or fields in a field theory) than just the classical
ones the effective constraints are imposed on all these
parameters. Additional variables include, e.g., the spread or
deformations of wave packets in addition to expectation values
identified with classical variables. If the quantum constraints
preserve the first-class nature of the classical constraints, one has
by definition $\{H_{\rm eff},D_{\rm eff}\}=\frac{1}{i\hbar}
\langle[\hat{H},\hat{D}]\rangle$ as a first-class set of
constraints. General covariance is thus preserved.

However, effective descriptions not only entail taking expectation
values but also a truncation of the infinitely many quantum
variables to a finite set (which means finitely many fields in a
field theory). This corresponds, in some sense, to the derivative
expansion done in effective actions to arrive at a finite sum of
local correction terms. In particular, in this paper we completely
ignored, as a first approximation, all quantum variables and
correction terms they imply. Such truncations usually lead to
effective constraints which do not exactly preserve covariance.
Nevertheless, in perturbative regimes of quantum corrections the
equations are consistent: One can choose a classically motivated
gauge, fixing $N$ and $N^a$, and compute corresponding
perturbative corrections as we did for the longitudinal gauge.
However, the gauge should not be too special as assumptions could
implicitly be used which will no longer hold with quantum
corrections. This would be the case if one used conditions on
spatial metric components to achieve a certain gauge, obtained by
solving gauge transformation equations for lapse and shift. Since
gauge transformations change themselves when constraints are
corrected, such a gauge would not be safe for the derivation of
corrected equations. Specifying lapse and shift directly, such as
$N^a=0$ in longitudinal gauge, is safer because it can be done in
the same way with any corrected constraints. This gauge is not
complete, but after having computed the corrected constraints one
can combine the remaining variables to gauge invariant quantities.

Non-perturbative regimes, such as those below the peaks of our
correction functions, have to be treated with more care since the
classical background geometry is strongly modified there. Usually,
additional quantum variables are required to describe the situation
and to discuss gauge issues fully. (The homogeneous background
evolution, on the other hand, is safe even in this regime since it is
subject to only one constraint. This will automatically commute with
itself, thus being first class.)

\section{Discussion}

We have presented in detail the Hamiltonian derivation of cosmological
perturbation equations for scalar modes in longitudinal gauge around a
flat isotropic background. The same scheme, of course, applies to
other gauges and also under inclusion of vector and tensor modes and
for perturbations around different backgrounds. As in the case of
scalar modes, due care has to be taken in deciding when a gauge or
mode selection is to be specified.

Since our main interest is to compute corrections from a canonical
quantization of gravity, typical such correction functions have been
included. We have seen how simple modifications of the constraints can
propagate to more involved corrections of equations of motion. We
emphasize that we have not presented a complete set of effective
equations including all possible correction terms. Alternative
corrections can arise, and moreover gauge issues have to be studied.

For applications, it is important to note that not only coefficients
in evolution equations are corrected, but also constraints are
modified. Since constraints generate gauge transformations, the form
of gauge invariant variables changes, too. For instance, it is not
sufficient to take the classical expression of the gauge invariant
curvature perturbation ${\cal R}=\psi + {\cal
H}\delta\varphi/\dot{\varphi}$ and use corrected equations of motion
for all variables involved. A complete treatment requires correction
terms in matter and metric equations as derived above, as well as in
expressions for the relevant quantities to be related to
observations. Ignoring any of the ingredients in general can lead to
misleading conclusions. Nevertheless, some qualitative conclusions can
be drawn. For instance, a modified evolution equation for ${\cal R}$
can be derived which implies correction terms leading to a slight
non-conservation of this curvature perturbation
\cite{InhomEvolve}. While the quantity ${\cal R}$
itself will have to be corrected as the relevant gauge-invariant
quantity, implying additional corrections to the evolution of
curvature perturbations, this is unlikely to happen in such a way that
all corrections from equations of motion and gauge invariance
properties conspire to cancel each other. For the precise form of
non-conservation, however, all these effects have to be taken into
account. Still, interesting qualitative effects for cosmological
phenomenology have already materialized.

We have started here a program to derive effects systematically and
presented a first set of corrected constraints as well as evolution
equations. A systematic study of different gauges and of observable
implications is still to be done. The derivation in a Hamiltonian
formulation as well as the use of Ashtekar variables are crucial for
the inclusion of effective quantum gravity effects in modified
perturbation equations if canonical quantum gravity, in particular
loop quantum gravity, is employed.

Primary dynamical objects are then the constraints, rather than
Lagrangians, which are modified by quantum effects. Without regarding
gravitational parts and all matter energy-momentum terms, changes to
the classical behavior in an inflationary context have been considered
in \cite{PowerLoop,ScaleInvLQC,CCPowerLoop} in a strongly modified
regime of background correction functions and in \cite{PowerPert} in a
perturbative regime. Ignoring corrections in gravitational parts of
the equations corresponds to choosing a flat gauge in which no metric
perturbations are present. The availability of this gauge choice is
based on classical reasoning, and has to be reconsidered with gauge
transformations generated by the quantum corrected constraints. Our
treatment is more general since we allowed metric perturbations $\psi$
and matter perturbations $\delta\varphi$ to be independent before they
are to be combined to a quantity gauge invariant under the quantum
corrected transformations. This allowed us to discuss non-conservation
of curvature perturbations as a new effect.

Following the lines of derivations in this paper, basic effects in the
constraints then translate unambiguously into effects in perturbation
equations. Since several of the variational equations have to be
combined in different manners, even simple modifications in the
constraints can have complex implications at the level of perturbation
equations. Modifications one expects on general grounds are regular
versions of any inverse power of metric variables such as those of
$p$ in the Hamiltonian constraint, the spin connection and the
matter Hamiltonian \cite{QSDV,Ambig,AmbigConstr,Spin,SphSymmHam},
higher order corrections as powers of $k$
\cite{SemiClass,DiscCorr,SemiClassEmerge} and higher derivative terms
in space as well as in time \cite{Karpacz}. All this gives rise to
characteristic correction functions which can be computed at least
qualitatively. In the perturbation equations, coefficients as well as
the derivative order of the equations can then change and differ
considerably from the classical ones in strong quantum regimes.

We have illustrated this throughout the paper with corrections which
are expected from inverse power modifications. Those corrections are
easiest to implement and to deal with because they change only
coefficients but not the type of perturbation equations. They are also
expected to be stronger in inhomogeneous situations
\cite{InhomLattice}. One expects four different correction functions,
two for the gravitational Hamiltonian and two for the matter
Hamiltonian. When they equal one, classical behavior is reproduced,
while on small scales they can differ considerably from one and lead
to modified and new coefficients. On very small scales, i.e.\ in
regimes where correction functions are not Taylor expandable around
the value one, cosmological perturbation theory is more difficult to
apply.

There are many effects from quantum gravity in combination, and even
different implementations depending on the quantization scheme used
for constraints. An effective analysis shows which of the terms are
most crucial for physical consequences and should be fixed. Other
corrections on which the behavior does not depend so sensitively can
then first be ignored. We can clearly see this from our example, where
the correction functions $\beta$ from the spin connection and $\sigma$
from the matter gradient term do not play as important roles as the
functions $\alpha$ and $D$. This is fortunate in particular for
$\beta$ because there is no tight prescription for its behavior in the
full theory. Also the function $\tilde{\beta}$, which could equal
$\beta$ or simply one depending on how one deals with the
diffeomorphism constraint, only appears once in the final equations
(\ref{PertMod3}) and in a way which does not significantly change the
behavior given by $\alpha$-corrections. (For large $\bar{p}$,
$1-\tilde{\beta}$ is negative while $-\alpha'\bar{p}/\alpha$ is
positive. Due to the perturbative form (\ref{alphapert}) of the
functions as a power series in $\ell_{\rm P}^2/\bar{p}$, however, the
correction from $\alpha$ is dominant in this regime and determines the
sign of the correction. For small $\bar{p}$, on the other hand,
effects from $\tilde{\beta}$ can be more pronounced but perturbation
theory is more complicated.)  The most sensitive corrections at the
level of linearized perturbations around flat space are thus those
coming from $\alpha$ and $D$. A phenomenological analysis then shows
which behavior of these functions is preferred.




\section*{Acknowledgements}

We thank Jim Lidsey, Roy Maartens, David Mulryne, Nelson Nunes and
Reza Tavakol for discussions especially about aspects of
cosmological perturbations. The work of MB was supported in part
by NSF grant PHY-05-54771. HH was supported by the fellowship
A/04/21572 of Deutscher Akademischer Austauschdienst (DAAD). MK
was supported by the Center for Gravitational Wave Physics under
NSF grant PHY-01-14375, PS was supported by NSF grants PHY-0354932
and PHY-04566913 and the Eberly research funds of Penn State.


\begin{thebibliography}{10}

\bibitem{CosmoPert}
V.~F.\ Mukhanov, H.~A.\ Feldman, and R.~H.\ Brandenberger,
\newblock Theory of cosmological perturbations,
\newblock {\em Phys.\ Rept.} 215 (1992) 203--333

\bibitem{Rov}
C.\ Rovelli,
\newblock {\em Quantum Gravity},
\newblock Cambridge University Press, Cambridge, UK, 2004

\bibitem{ThomasRev}
T.\ Thiemann,
\newblock Introduction to Modern Canonical Quantum General Relativity,
  [gr-qc/0110034]

\bibitem{ALRev}
A.\ Ashtekar and J.\ Lewandowski,
\newblock Background independent quantum gravity: A status report,
\newblock {\em Class.\ Quantum Grav.} 21 (2004) R53--R152, [gr-qc/0404018]

\bibitem{LivRev}
M.\ Bojowald,
\newblock Loop Quantum Cosmology,
\newblock {\em Living Rev.\ Relativity} 8 (2005) 11, [gr-qc/0601085],
\newblock {\tt http://relativity.livingreviews.org/Articles/lrr-2005-11/}

\bibitem{AshVar}
A.\ Ashtekar,
\newblock New Hamiltonian Formulation of General Relativity,
\newblock {\em Phys.\ Rev.\ D} 36 (1987) 1587--1602

\bibitem{AshVarReell}
J.~F.\ Barbero~G.,
\newblock Real Ashtekar Variables for Lorentzian Signature Space-Times,
\newblock {\em Phys.\ Rev.\ D} 51 (1995) 5507--5510, [gr-qc/9410014]

\bibitem{ADM}
R.\ Arnowitt, S.\ Deser, and C.~W.\ Misner,
\newblock {\em The Dynamics of General Relativity},
\newblock Wiley, New York, 1962

\bibitem{Immirzi}
G.\ Immirzi,
\newblock Real and Complex Connections for Canonical Gravity,
\newblock {\em Class.\ Quantum Grav.} 14 (1997) L177--L181

\bibitem{IsoCosmo}
M.\ Bojowald,
\newblock Isotropic Loop Quantum Cosmology,
\newblock {\em Class.\ Quantum Grav.} 19 (2002) 2717--2741, [gr-qc/0202077]

\bibitem{SphSymm}
M.\ Bojowald,
\newblock Spherically Symmetric Quantum Geometry: States and Basic Operators,
\newblock {\em Class.\ Quantum Grav.} 21 (2004) 3733--3753, [gr-qc/0407017]

\bibitem{InhomLattice}
M.\ Bojowald,
\newblock Loop quantum cosmology and inhomogeneities, {\em Gen.\ Rel.\ Grav.}
to appear, [gr-qc/0609034]

\bibitem{QuantCorrPert}
M.\ Bojowald, H.\ Hern\'andez, M.\ Kagan, P.\ Singh, and A.\
Skirzewski,
\newblock Effective constraints of loop quantum gravity, to appear

\bibitem{Spin}
M.\ Bojowald, G.\ Date, and K.\ Vandersloot,
\newblock Homogeneous loop quantum cosmology: The role of the spin connection,
\newblock {\em Class.\ Quantum Grav.} 21 (2004) 1253--1278, [gr-qc/0311004]

\bibitem{SphSymmHam}
M.\ Bojowald and R.\ Swiderski,
\newblock Spherically Symmetric Quantum Geometry: Hamiltonian Constraint,
\newblock {\em Class.\ Quantum Grav.} 23 (2006) 2129--2154, [gr-qc/0511108]

\bibitem{QSDI}
T.\ Thiemann,
\newblock \noopsort{QSD Ia}Quantum Spin Dynamics {(QSD)},
\newblock {\em Class.\ Quantum Grav.} 15 (1998) 839--873, [gr-qc/9606089]

\bibitem{Ambig}
M.\ Bojowald,
\newblock Quantization ambiguities in isotropic quantum geometry,
\newblock {\em Class.\ Quantum Grav.} 19 (2002) 5113--5130, [gr-qc/0206053]

\bibitem{ICGC}
M.\ Bojowald,
\newblock Loop Quantum Cosmology: Recent Progress,
\newblock In {\em Proceedings of the International Conference on Gravitation
  and Cosmology (ICGC 2004), Cochin, India},
  {\em Pramana} 63 (2004) 765--776, [gr-qc/0402053]

\bibitem{BoundCoh}
J.\ Brunnemann and T.\ Thiemann,
\newblock On (Cosmological) Singularity Avoidance in Loop Quantum Gravity,
\newblock {\em Class.\ Quantum Grav.} 23 (2006) 1395--1427, [gr-qc/0505032]

\bibitem{DegFull}
M.\ Bojowald,
\newblock Degenerate Configurations, Singularities and the Non-Abelian Nature
  of Loop Quantum Gravity,
\newblock {\em Class.\ Quantum Grav.} 23 (2006) 987--1008, [gr-qc/0508118]

\bibitem{ALMMT}
A.\ Ashtekar, J.\ Lewandowski, D.\ Marolf, J.\ Mour\~ao, and T.\
Thiemann,
\newblock Quantization of Diffeomorphism Invariant Theories of Connections with
  Local Degrees of Freedom,
\newblock {\em J.\ Math.\ Phys.} 36 (1995) 6456--6493, [gr-qc/9504018]

\bibitem{Karpacz}
M.\ Bojowald and A.\ Skirzewski,
\newblock Quantum Gravity and Higher Curvature Actions,
\newblock In {\em Current Mathematical Topics in Gravitation and Cosmology
  (42nd Karpacz Winter School of Theoretical Physics)}, [hep-th/0606232]

\bibitem{InhomEvolve}
M.\ Bojowald, H.\ Hern\'andez, M.\ Kagan, P.\ Singh, and A.\
Skirzewski,
\newblock Formation and evolution of structure in loop cosmology, to appear

\bibitem{PowerLoop}
G.~M.\ Hossain,
\newblock Primordial Density Perturbation in Effective Loop Quantum Cosmology,
\newblock {\em Class.\ Quantum Grav.} 22 (2005) 2511--2532, [gr-qc/0411012]

\bibitem{ScaleInvLQC}
D.~J.\ Mulryne and N.~J.\ Nunes,
\newblock Constraints on a scale invariant power spectrum from superinflation
  in LQC, [astro-ph/0607037]

\bibitem{CCPowerLoop}
G.\ Calcagni and M.~V.\ Cort\^es,
\newblock Inflationary scalar spectrum in loop quantum cosmology,
  [gr-qc/0607059]

\bibitem{PowerPert}
S.\ Hofmann and O.\ Winkler,
\newblock The Spectrum of Fluctuations in Inflationary Quantum Cosmology,
  [astro-ph/0411124]

\bibitem{QSDV}
T.\ Thiemann,
\newblock {QSD V}: Quantum Gravity as the Natural Regulator of Matter Quantum
  Field Theories,
\newblock {\em Class.\ Quantum Grav.} 15 (1998) 1281--1314, [gr-qc/9705019]

\bibitem{AmbigConstr}
K.\ Vandersloot,
\newblock On the Hamiltonian Constraint of Loop Quantum Cosmology,
\newblock {\em Phys.\ Rev.\ D} 71 (2005) 103506, [gr-qc/0502082]

\bibitem{SemiClass}
M.\ Bojowald,
\newblock The Semiclassical Limit of Loop Quantum Cosmology,
\newblock {\em Class.\ Quantum Grav.} 18 (2001) L109--L116, [gr-qc/0105113]

\bibitem{DiscCorr}
K.\ Banerjee and G.\ Date,
\newblock Discreteness Corrections to the Effective Hamiltonian of Isotropic
  Loop Quantum Cosmology,
\newblock {\em Class.\ Quant.\ Grav.} 22 (2005) 2017--2033, [gr-qc/0501102]

\bibitem{SemiClassEmerge}
P.\ Singh and K.\ Vandersloot,
\newblock Semi-classical States, Effective Dynamics and Classical Emergence in
  Loop Quantum Cosmology,
\newblock {\em Phys.\ Rev.\ D} 72 (2005) 084004, [gr-qc/0507029]

\end{thebibliography}
\newcommand{\noopsort}[1]{}

\end{document}